# Test of CZT Detectors with Different Pixel Pitches and Thicknesses

Qiang Li, Alfred III Garson, Ira Jung, Michael Groza, Paul Dowkontt, Richard Bose, Garry Simburger,

Arnold Burger, Henric Krawczynski

*Abstract*--The Modified Horizontal Bridgman (MHB) process produces Cadmium Zinc Telluride (CZT) crystals with high yield and excellent homogeneity. Various groups, including our own, previously reported on the test of 2×2×0.5 cm$^3$ MHB CZT detectors grown by the company Orbotech and read out with 8×8 pixels. In this contribution, we describe the optimization of the photolithographic process used for contacting the CZT detector with pixel contacts. The optimized process gives a high yield of good pixels down to pixel diameters/pitches of 50 microns. Furthermore, we discuss the performance of 0.5 cm and 0.75 cm thick detectors contacted with 64 and 225 pixel read out with the RENA-3 ASICs from the company NOVA R&D.

## I. MOTIVATION FOR FINELY PIXILATED CZT DETECTORS: ASTROPARTICLE PHYSICS EXPERIMENTS

THE compound semiconductor cadmium zinc telluride (CZT) has become the detector material of choice for many applications owing to its large bandgap, excellent spatial and energy resolution at room temperature, high stopping power, and high photo effect cross section [1]. CZT detectors find application in medical imaging and homeland security applications (e.g. [2,3]), and are used for a number of astroparticle physics experiments.

[1] The Swift satellite [4] launched in 2004 uses an array of 32,768 CZT detectors each with a size of 4×4×2 mm$^3$. The instrument fulfills the high expectations and has made a number of exciting discoveries concerning the hosts of long GRBs (star forming regions) and short GRBs (regions with little star-formation), the discovery of a complex afterglow phenomenology including powerful late afterbursts, and the detection of GRB 050904 at a high redshift of z=6.29 (see [5] for a recent review). The proposed EXIST mission [6] (Energetic X-ray Imaging Survey Telescope) design of

NASA's future Black Hole Finder Probe plans to use between 5 m$^2$ -6 m$^2$ of 0.5 cm thick CZT detectors. The EXIST mission would detect GRBs at high (~10) redshifts, and would be able to detect the X-ray emission from supermassive black holes up to redshifts between 1 and 2.

An example of a particle physics experiment is the neutrinoless double beta decay experiment COBRA [7]. A future large-scale version of COBRA would comprise ~400 kg of CZT detectors fabricated with Cd enriched to > 90% in the double beta emitter $^{116}$Cd. The detection of neutrino-less double beta decays would establish that neutrinos are Majorana particles. The measurement of the neutrino-less double beta decay rate would constrain the masses of the neutrino mass eigenstates.

The EXIST mission would require pixilated CZT detectors with a pixel pitch of either 0.6 mm or 1.25 mm. The large-scale COBRA experiment would either use coplanar grid detectors, or pixelated detectors with a pixel pitch of 200 microns. The fine pixelization would make it possible to track the electrons from the double beta decays [8].

With the aim to optimize detectors for EXIST and for COBRA, we developed an optimized photolithographic process to realize pixels with small pixel pitches. The results of the optimization will be described in Sect. 2. The performance achieved with 64 and 225 pixel detectors read out with the RENA-3 ASIC will be discussed in Sect. 3. Section 4 gives a summary and an outlook.

## II Optimized Photolithographic Process

In the following we show results obtained with CZT detectors from the company Orbotech Medical Solutions LTd [9]. Orbotech uses the Modified Horizontal Bridgman (MHB) process to grow the CZT. The process results in a high yield and excellent homogeneity. Earlier studies of Orbotech detectors used 2×2×0.5cm$^3$ substrates with 64 pixel (pixel pitch: 2.5 mm) (e.g. [10,11]). The thick detectors fabricated so far have pixel pitches of d ~1 mm, and thus achieve spatial resolution of $2.35/\sqrt{12} \approx 0.7$ mm FWHM (full width half maximum). We explore new territory by fabricating and testing the pixel thick CZT detectors with pitches in the sub-mm regime.

We contact the detectors in a class-100 cleanroom dedicated to the fabrication of CZT detectors. The detectors are first polished with different grades of abrasive paper, and alumina suspension with particles sizes down to 0.05 µm. We found that etching with a Br-Methanol solution does not only improve the electrical properties of the contacts, but also improved the adhesion of the contacts. After surface preparation, a standard photolithographic process is used consisting of photoresist application, pre-baking, exposure, post-baking, and development. The contacts are then evaporated with an electron beam evaporator, and the remaining photoresist and metal films are removed with acetone. We systematically optimized the parameters of the photolithographic process. For this purpose, a photomask with a number of different chessboard patterns with square widths of between 5 microns and 2.5 mm. We used an optical microscope to determine the yield of good pixels for each combination of processing parameters.

For a few exemplary tests, Fig. 1 shows the fraction of good pixels as function of square width. We obtain a ~100% yield of good pixels for chessboard patterns with square widths down to 100 microns. For 50 microns, a fraction of the 2×2 cm$^2$ detector surface has a high yield of good pixels, but the pixels do not adhere properly for some surface areas.

Fig. 2(a) shows a 2×2×0.5 cm$^3$ detector from our fabrication with 225 (15×15) anode pixel and an Au cathode. The pixels have a diameter of ~1mm and the pixel pitch is 1.27mm. Fig. 2(b) shows a chessboard pattern with a square width of 50 microns.

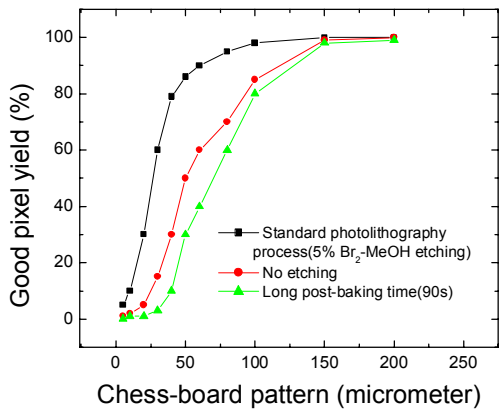

(a)

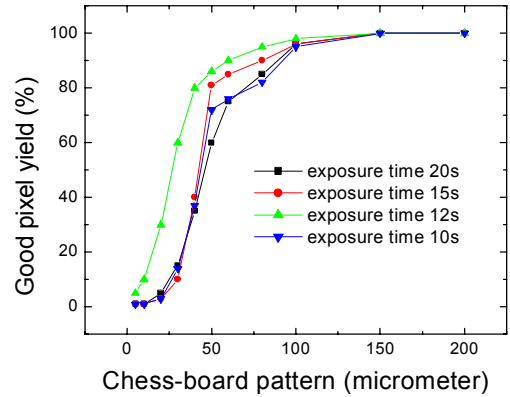

(b)

Fig. 1 The fraction of good pixels as function of pixel pitch for different process parameters.

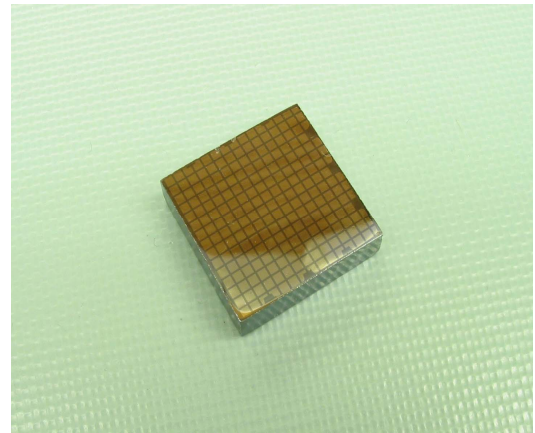

(a)

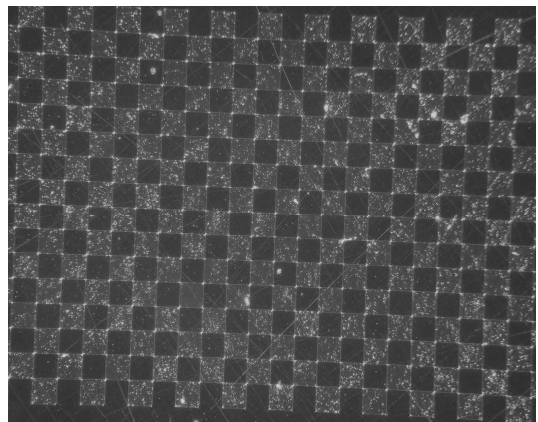

(b)

Fig. 2 The left panel (a) shows an 2×2×0.5cm$^3$ Orbotech detector contacted with 15x15 Ti anode pixels and an Au cathode. The pixels have a diameter of ~1mm and the pixel pitch is 1.27mm. The right panel (b) shows a CZT detector contacted with a chess-board pattern with 50 micron diameter squares.

# III Energy Spectra Obtained with the RENA-3 Readout System

## A. Experimental Setup

Previously we reported on the tests of CZT detectors with discrete Amptek A250 amplifiers and with a 16-channel ASIC [10]. These two systems had electronic noise corresponding to ~1% FWHM at 662 keV. Here we report on first results obtained with the RENA-3 ASICs from the company NOVA R&D [12]. We use an ASIC evaluation system developed by NOVA R&D. We mount the detectors in a Delrin holder on custom built PC boards. The detectors are contacted with gold plated pogo-pins. The bias is applied to the cathode. Fig. 3 shows the RENA-3 evaluation system loaded with a $2\times2\times0.5cm^3$ Orbotech detector.

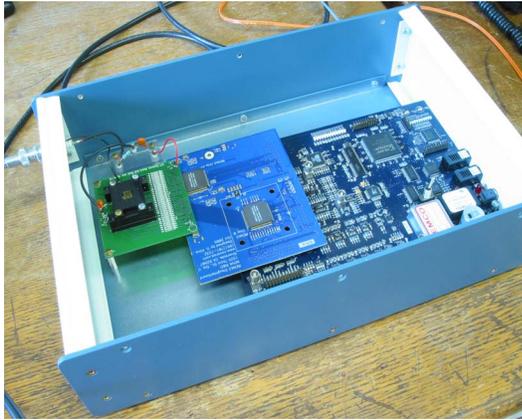

Fig.3: The panel shows a photograph of the NOVA R&D evaluation system. The system uses 2 ASICs to read out the pixels and the cathode of a $2\times2\times0.5cm^3$ detector. The pixelated detectors are mounted in a Delrin holder, and contacted with spring-loaded pogo-pins that press a "ZEBRA" pad (a rubber pad with vertical wire bundles) against the detector pixels.

## B. $^{137}$Cs source (662 keV), energy spectra test with 64 pixel.

Fig. 4 shows the results from a test of a $2\times2\times0.5cm^3$ Orbotech detector with a 662 keV $^{137}$Cs source (-500 V bias). The detector was contacted with 64 pixels, but only 60 of the 64 pixels were connected to the readout system. For this specific detector, we obtained good energy spectra for 59 out of 60 (98%) pixels (Fig. 4(a)). The ASIC was used to read out the anode pixels and the cathode, and Fig. 4(b) shows the pixel-to-cathode correlation. This correlation is later used for correcting the anode pixels for the depth of interaction (DOI). For one pixel, Fig. 5(a) shows the anode to cathode correlation. In Fig. 5(b) and (c) the energy spectra before and after correction of the anode charge with the anode-to-cathode charge ratio are shown. The FWHM energy resolutions are 6.1% and 3.2% and the peak-to-valley ratios are 7.3 and 15.2 before and after DOI correction, respectively. With our other readout systems, we get somewhat better results for the same detector (2.5% FWHM). We scrutinized the data for events with multiple pixel hits. Such events are either charge sharing events, or Compton events. We selected two-pixel events by requesting that two adjacent pixels show a signal exceeding 25% of the maximum possible signal (662 keV energy deposition right below a single pixel). Summing the amplitudes from both pixels, and correcting the signal for the DOI with the help of the ratio of the summed signal divided by the cathode signal, we obtain the results shown in Fig. 6. Interestingly, the two-pixel events produce 5% less charge than the one-pixel events. A possible explanation is charge loss caused by charge drifting to the gap between pixels, and being trapped at the surface for sufficiently long time to evade detection.

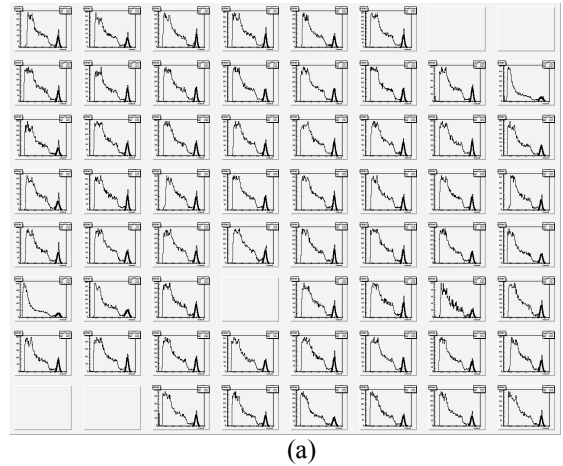

(a)

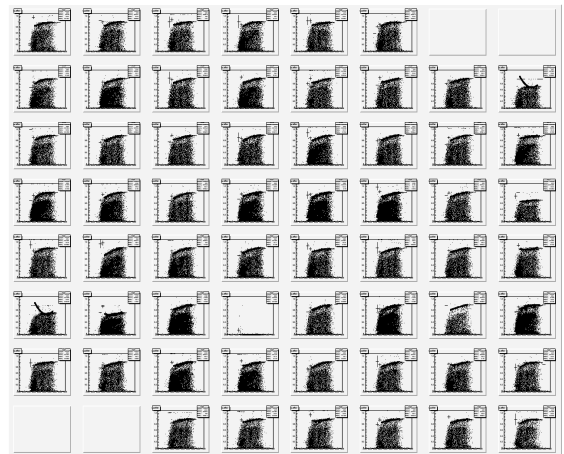

(b)

Fig. 4 662 keV($^{137}$Cs source) energy spectra of 60 pixel on a $2\times2\times0.5$ cm$^3$ CZT detector (a) and the cathode-to-anode signal correlation for the same pixels (b). 59 out of 60 pixels showed proper signals.

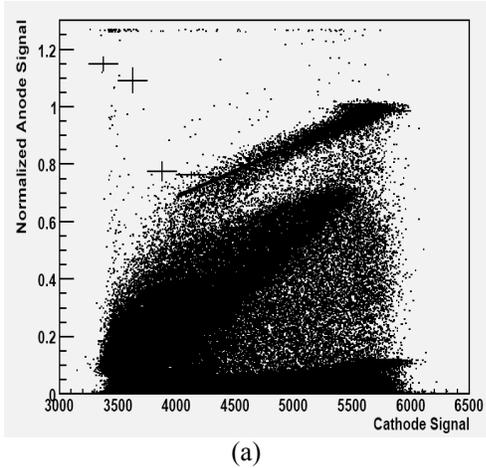

(a)

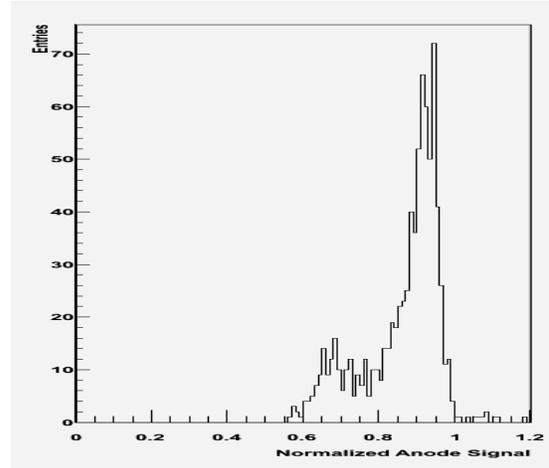

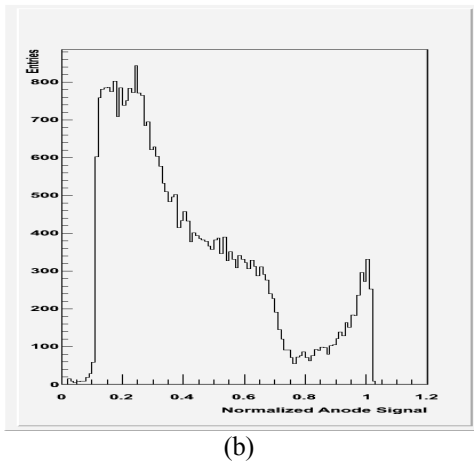

(b)

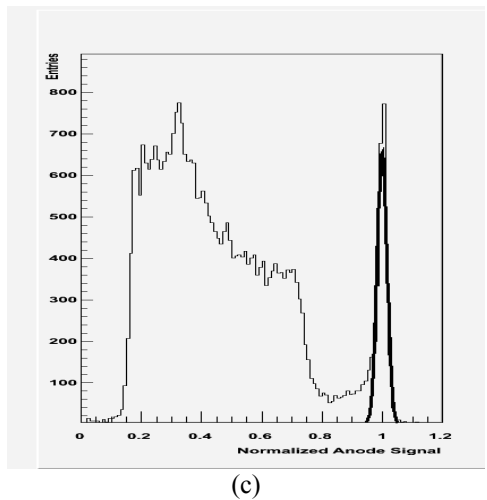

(c)

Fig. 5 For the same CZT detector as in Fig. 4, this figure shows the results of a single pixel ($^{137}$Cs source, 662 keV). Panel (a) shows the anode-to-cathode correlation. Panel (b) shows the "raw" pixel spectrum, and panel (c) shows the pixel spectrum after correcting for the Depth of Interaction (DOI) with the help of the anode-to-cathode ratio. The DOI correction improves the energy resolution from 6.1% FWHM to 3.2% FWHM.

Fig. 6 For the same CZT detector as in Fig. 4, this figure shows the $^{137}$Cs energy spectra of two pixel events (induced charge exceeding 25% of the maximum induced charge for both pixels). The energy spectrum was obtained by summing the signals from the two pixels, and correcting the summed signal for the Depth of Interaction with the help of the pixel-to-cathode correlation of single-pixel events.

C. $^{137}$Cs source (662 keV), thicker detectors and smaller pixel pitches

Fig. 7(a) shows the $^{137}$Cs energy spectra obtained with a 0.75 cm thick detector (volume: 2×2×0.75 cm$^3$) contacted with 8×8 In pixels and an In cathode. The 662 keV photopeak has a FWHM width of 2.2% FWHM. Fig. 7(b) shows the energy spectrum for a 0.5 cm thick detector contacted with 15 x15 Ti pixel and an Au cathode.

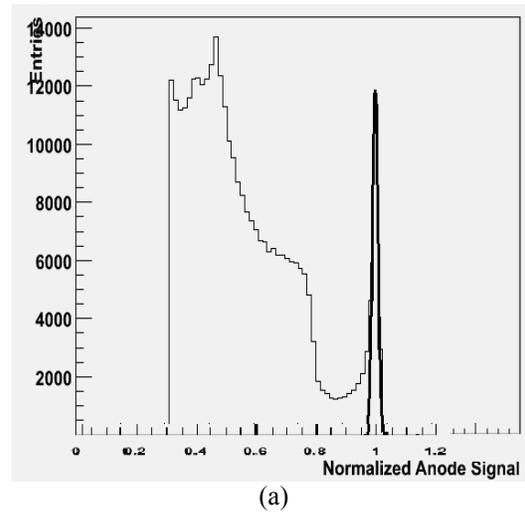

(a)

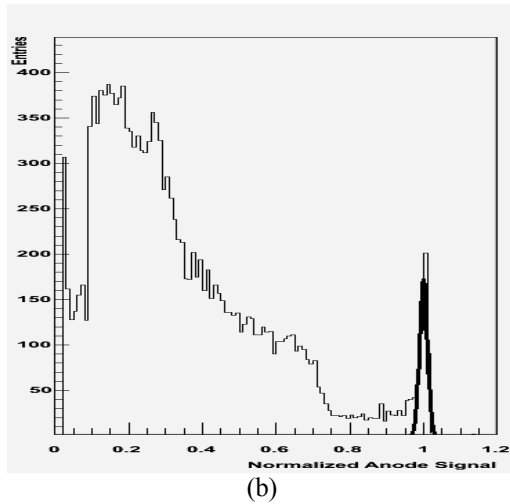
(b)

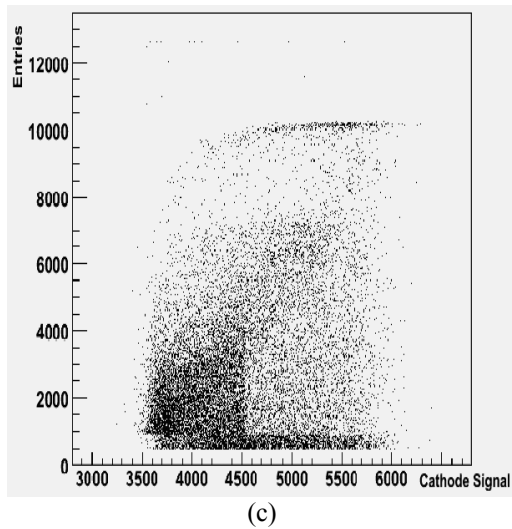
(c)

Fig. 7 Panel (a) shows a $^{137}$Cs (662 keV) energy spectrum obtained with a 0.75 cm thick Orbotech detector (8×8 In pixels, volume: 2×2×0.75 cm$^3$, energy resolution: 2.2% FWHM). Panel (b) shows an $^{137}$Cs energy spectrum obtained with a 0.5 cm thick detector contacted with 15x15 Ti pixels (pitch: 1.25 mm, volume: 2×2×0.5 cm$^3$, energy resolution 1.6% FWHM). Panel (c) shows for the same detector the anode-to-cathode correlation.

The 225 pixel detector achieves 662 keV energy resolutions of 1.6% FWHM. Fig. 7(c) shows the anode-to-cathode correlation for the same detector. It can be recognized that the DOI correction is much less important for this detector with a pixel pitch of 1.25 mm than for the detector with 2.5 mm pixel pitch (compare Fig. 5(a)).

## IV Summary and Outlook

After optimizing the photolithography process parameters, we can fabricate pixelated CZT detectors with pixel pitches between 50 and 100 microns. For pitches of 100 microns or more, we obtain near 100% yield of good pixels. In this contribution we have shown the results obtained with detectors with 2.5 mm and 1.25 mm pixel pitch. The detector with 1.25 mm pixel pitch showed a better 662 keV energy resolution (1.6% FWHM) than the detector with 2.5 mm pitch (3.2% FWHM). We started to use thicker CZT substrates. A first 0.75 cm thick detector achieves a 662 keV energy resolution of 2.2% FWHM. All these results were obtained with readout system based on the RENA-3 ASIC. Future work will focus on (a) further reducing the readout noise of the RENA-3 based readout system, (b) test of alternative ASICs, (c) fabrication and test of detectors with pixel pitches between 0.2 and 1.25 mm, (d) fabrication and test of Orbotech MHB detectors with thicknesses between 0.2 cm and 1 cm.

ACKNOWLEDGEMENTS

This work is supported by NASA under contract NNX07AH37G, and the DHS under contract 2007DN077ER0002.

REFERENCES

[1] R. B. James, B. Brunett, J. Heffelfinger, J. Van Scyoc, J. Lund, F. P. Doty, C. L. Lingren, R. Olsen, E. Cross, H. Hermon, H. Yoon, N. Hilton, M. Schieber, E. Y. Lee, J. Toney, T. E. Schlesinger, M. Goorsky, W. Yao, H. Chen and A. Burger, Material properties of large-volume cadmium zinc telluride crystals and their relationship to nuclear detector performance, J. Electron. Mater. , 27 (1998), 788-799.
[2] Loïck Verger, Eric Gros d'Aillon, Olivier Monnet, Guillaume Montémont and Bernard Pelliciari, New trends in γ-ray imaging with CdZnTe/CdTe at CEA-Leti, Nucl. Instru. Meth. In Phys. A, 571, (2007) 33-43.
[3] O. Limousin, New trends in CdTe and CdZnTe detectors for X- and gamma-ray applications, Nucl. Instru. Meth. In Phys. A, 504, (2003) 24-37.
[4] N. Gehrels, et al., Astrophys. J., 611, 1005 (2004).
[5] N. Gehrels, J. K. Cannizzo, J. P. Norris, NJPh, 9, 37 (2007).
[6] J. E. Grindlay, and The Exist Team, AIP 836, 631 (2006).
[7] K. Zuber, Physics Letters B, 519, 1-7 (2001).
[8] T. R. Bloxham, and M. Freer, 572, 722 (2007).
[9] Orbotech Medical Solutions LTd., 10 Plaut St., Park Rabin, P.O.Box: 2489, Rehovot, Israel, 76124.
[10] Jung, I., Garson, A. III, Krawczynski, H., Burger, A., Groza, M., Detailed studies of pixelated CZT detectors grown with the modified horizontal Bridgman method, Astroparticle Physics, 28, (2007) 397-408 [arXiv: 0710.2655].
[11] Hong, J., et al., SPIE Conference Proceedings "Hard X-Ray and Gamma-Ray Detector Physics IX", 6706-10 (2007), in press [arXiv:0709.2719].
[12] NOVA R&D, Inc., 1525 Third Street, Suite C, Riverside, CA 92507-3429, USA.